\documentclass[prb,showpacs,twocolumn,aps,a4paper,amsmath,amssymb]{revtex4}

\usepackage{dcolumn}
\usepackage{amsmath}
\usepackage{graphicx}
\usepackage{latexsym}
\usepackage{amsfonts}
\usepackage{amssymb}
\usepackage{bm}

\DeclareGraphicsExtensions{.pdf,.gif,.jpg}

\def\a{\alpha}
\def\b{\beta}

\def\G{\Gamma}

\def\s{\sigma}

\def\e{\epsilon}

\def\D{\Delta}

\def\bra{\langle}
\def\ket{\rangle}

\def\ua{\uparrow}
\def\da{\downarrow}

\newcommand{\be}{\begin{equation}}
\newcommand{\ee}{\end{equation}}
\newcommand{\beq}{\begin{eqnarray}}
\newcommand{\eeq}{\end{eqnarray}}

\begin{document}

\title{Renormalization of the Coulomb blockade gap due to extended tunneling 
in nanoscopic junctions}

\author{E. Perfetto}
\affiliation{Dipartimento di Fisica, Universit\`{a} di Roma 
Tor Vergata, Via della Ricerca Scientifica 1, I-00133 Rome, Italy}

\begin{abstract}
In this work we discuss the combined effects of finite-range 
electron-electron interaction and finite-range tunneling on the transport 
properties of ultrasmall tunnel junctions.
We show that the Coulomb blockade  phenomenon is deeply influenced 
by the interplay between the geometry and the screening properties of 
the contacts. In particular if the 
interaction range is smaller than the size of the tunneling region 
a ``weakly correlated'' regime emerges 
in which the Coulomb blockade gap $\D$ is significantly reduced.
In this regime $\D$ is not simply given
by the conventional charging energy of the junction, since it is
strongly renormalized by the energy that  
electrons need to tunnel over the extended contact.
\end{abstract}


\maketitle

\section{Introduction}
\label{sec1}

The transport properties of nanoscale systems are strongly 
affected by electron-electron interactions that may cause large deviations
from the Ohm's law.\cite{averin,grabert}
When two conductors are connected by a tunnel junction with 
capacitance $C$, electrostatic 
effects inhibit  the current flow for applied voltages $V 
<1/2C$.\cite{units}
This phenomenon is known as Coulomb blockade (CB), and is at the origin of
the observed gap  around $V = 0$ in the $I$-$V$ curve of
a variety of 
systems.\cite{exp1,exp2,exp3,exp4,exp5,exp6,exp7,exp8,exp9,exp10,exp11,exp12}
The widely accepted dynamical theory of the CB\cite{theor1,theor2} is based on the 
notion that the fluctuations generated by the thermal agitation of the charge
carriers inside the leads (Nyquist-Johnson 
noise) render the tunneling processes inelastic.
As a consequence the environment represents a 
frequency-dependent impedance, capable to adsorb energy from the 
tunneling electrons. Thus in the subgap region the effective
voltage felt by the electrons is drastically
reduced, resulting in a suppression of the current 
according to a power law with nonuniversal exponent.
At larger bias, however, the Ohmic regime is recovered, with the
$I$-$V$ curve having an offset of order $1/2C$.

As pointed out by some 
authors,\cite{sassetti1,sassetti2,steiner,sonin,safi} the peculiar 
power-law behavior of the tunneling current reveals an interesting relationship 
between the dynamical CB and the zero-bias anomaly predicted within 
the Luttinger liquid (LL) theory.\cite{kane,zba1,zba2}
In Ref. \onlinecite{sonin} it was noticed that the similar predictions of the 
two approaches arise from a close analogy
in the description of the tunneling processes.
In the semiclassical theory,\cite{theor1,theor2} the influence of the environment
is incorporated in a modification of the tunneling Hamiltonian that 
accounts for quantum fluctuations in the phase-difference
between the left and right side of the junction, which is formally 
equivalent to the tunneling term of a LL with barrier in the bosonized 
form.\cite{sonin}
This similarity has been further exploited by Safi and Saleur, who 
established a rigorous mapping between the one-channel coherent 
conductor in series with a resistance $R$ and the impurity problem in 
a LL.\cite{safi}
They found that the LL parameter $K$ of the 
effective interacting 
theory can be expressed as $K=(1+R/R_{0})^{-1}$, where $R_{0}$ is the 
resistance quantum.
This equivalence, however, holds only in the power-law regime and 
does not help to predict the 
magnitude of the CB gap $\D$, and neither provides a  
microscopic explanation of why 
the shifted Ohmic (SO) regime is recovered at large 
voltage. Sassetti et al. addressed these issues by showing that a 
finite-range interaction $U(x)$ within the LL model  is needed 
to describe the crossover from the power-law to the SO
behavior in the $I$-$V$ curve, where the the CB gap
takes the value $\D \sim 2U(0)$.\cite{sassetti1,sassetti2}

\begin{figure}[t]
\includegraphics[width=7.cm]{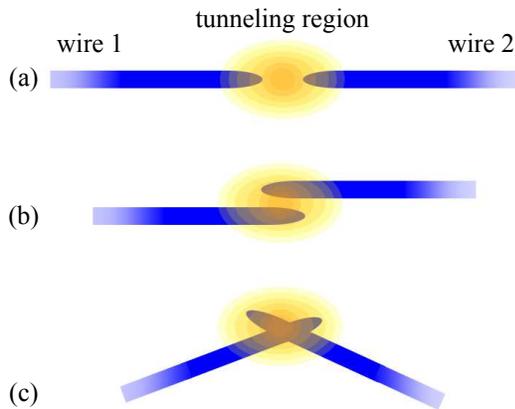}
\caption{Shematic representation of three possible extended contact 
geometries.}
\label{fig1}
\end{figure}

We would like to point out that the above results are valid under the 
assumption that the tunneling between the two conductors occurs only 
at their edges. However, in practice, due to the geometry of the 
junction (see e.g. Fig. \ref{fig1}) and due to the nontrivial (i.e. 
exponential) spatial 
dependence of the tunneling amplitude, the tunneling processes take place 
over a finite  
region.\cite{exttunn1,exttunn2,exttunn3,exttunn4,exttunn5,exttunn6}
Furthermore in these systems the size of the tunneling region and the
screening length are often comparable,\cite{vignale} and hence 
it is desirable to include and treat their effects on the same footing.

In this paper we study the transport properties of two semi-infinite 
wires with finite-range electron-electron interaction, linked via an 
extended contact close to the interface, as depicted in Fig. \ref{fig1}.
The wires are described within the open-boundary Tomonaga-Luttinger 
model, and the tunneling Hamiltonian is treated to linear order.
We show that if the interaction range is sufficiently small, 
the competition between screening and extended tunneling 
(ET) gives rise to a novel ``weakly correlated'' regime in which the
CB (i.e. power-law) regime and the SO regime  are
separated by an intermediate region characterized by a different power-law.
Remarkably this competition produces a sizable reduction of the CB
gap from the expected value $\D \sim 2U(0)$ to the renormalized value 
$\D \sim v/r$, $r$ being the 
extension of the contact region and $v$  
the velocity of the interacting quasiparticles.
This finding suggests that under certain conditions the CB gap 
is not directly related to the conventional charging energy of the junction,
but is strongly renormalized by the energy that
electrons need to tunnel over a region of extended size.

The plan of the paper is the following.
In the next Section we introduce the model and describe the general 
framework to calculate the $I$-$V$ characteristics of the junction.
Sections \ref{sec3}-\ref{sec6} are devoted to discuss different cases 
in which the screening length can be smaller or larger than the 
tunneling length. In Section \ref{sec7} we complete the analysis by 
introducing the spin.
Finally the summary and the main conclusions are drawn in Section 
\ref{sec8}.

\section{Model and formalism}
\label{sec2}

We start by considering the one-dimensional (1D) tunnel junction for 
spinless electrons with Hamiltonian
\be
H=H_{1}+H_{2}+H_{T}+H_{V},
\ee
where $H_{j}$ ($j=1,2$) describes the semi-infinite  
interacting wire $j$, $H_{T}$ the tunnel junction, and $H_{V}$ the 
applied bias voltage.
The wires with interaction potential $U$ are modeled as 
open-boundary Tomonaga-Luttinger liquids
according to\cite{units}
\beq
H_{j}&=& \frac{1}{2}\sum_{\alpha=R,L} [-2i \e_{\a} v_{F}\int_{0}^{\infty}dx 
\,
\psi^{\dagger}_{j\alpha}(x)\partial_{x} \psi_{j\alpha}(x) \nonumber \\
&+& \frac{1}{2}\int_{0}^{\infty}dx \, dy \, 
  U(|x-y|)\rho_{j}(x) \rho_{j}(y) \, ],
\label{model}
\eeq
where $\a$ denotes the chirality of the electrons with Fermi velocity 
$\e_{\a} v_{F}$ ($\e_{R/L}=\pm 1$), $\psi^{(\dagger)}_{j\alpha}$ is 
the annihilation (creation) operator of one electron in wire $j$ and 
chirality $\alpha$ with
density $\rho_{j}=\sum_{\a}:\psi^{\dagger}_{j\alpha} \psi_{j\alpha}:$, 
``$:\;\;:$'' being the normal ordering.
The junction between the two wires is modeled by the tunneling 
Hamiltonian
\be
H_{T}=\sum_{\a,\a'}\int_{0}^{\infty} dx\, dy\, \G(x,y)  
\psi^{\dagger}_{1\alpha}(x) \psi_{2\alpha'}(y) +\mathrm{h.c.},
\label{tunnham}
\ee
where the function $\G$ can eventually account 
for tunneling of electrons located not only at the boundaries
of the wires. In the above expression it is 
understood that the tunneling amplitude $\G$ depends on the distance
between one point at position 
$x$ in wire $1$ and another point at position $y$ in wire 
$2$.

For latter purposes it is convenient to 
write  $\G(x,y)=\G_{0}g(x,y)$ with $g$
an adimensional function encoding all the spatial dependence.
The junction is driven out of equilibrium by an external voltage
given by
\be
H_{V}=\sum_{j} V_{j} \int dx 
\rho_{j}(x)=\sum_{j}V_{j} N_{j},
\ee
with $N_{j}=\sum_{\a}N_{\a j}$ the number of electrons in the wire 
$j$ and $V=V_{1}-V_{2}$ the total applied bias.
The crucial quantity we are interested in is the tunneling current 
whose operator reads
\beq
J&=& \frac{dN_{1}}{dt}=-\frac{dN_{2}}{dt}\nonumber \\
&=&i\sum_{\a,\a'}\int_{0}^{\infty} dx\, dy\, \G(x,y)  
\psi^{\dagger}_{1\alpha}(x) \psi_{2\alpha'}(y) +\mathrm{h.c.} 
\eeq
Since we focus on the tunneling regime, in this work we 
evaluate the current to the second 
order in $\G_{0}$.
By employing the gauge transformation\cite{feldman,perfetto} $\psi_{j\a}\to 
\psi_{j\a}e^{iV_{j}t}$ 
the steady-state average current $I=\bra J \ket$ is (at zero 
temperature) 
\beq
I &=& \frac{1}{4}
\int_{-\infty}^{\infty}dt \,  e^{iVt}  \bra \Psi_{0}| [H_{T}(t),J(0)] 
|\Psi_{0}\ket \nonumber \\
&\equiv& \left(\frac{2 \G_{0} }{\pi a} \right)^{2}
\int_{0}^{\infty} dx_{1}\ldots dx_{4} \, g(x_{12}) 
g(x_{34}) \nonumber \\
&\times& \int_{-\infty}^{\infty}dt  \,  
e^{iVt}e^{-W(t,\{x_{i}\})},
\label{steady}
\eeq
where $|\Psi_{0}\ket$ is the interacting ground-state of the equilibrium 
uncontacted Hamiltonian $H_{1}+H_{2}$, and $H_{T}(t)$ and $J(0)$ are 
in Heisenberg representation with respect to $H_{1}+H_{2}$.
In the above equation we have used the short-hand notation 
$x_{i,j}=(x_{i},x_{j})$ and $\{ x_{i}\}=(x_{1},x_{2},x_{3},x_{4})$.
The function $W$ is the equilibrium phase correlation function and 
its Fourier transform is related to the 
probability $P(E)$ for a tunneling electron of exchanging the energy $E$
with the bath of interacting electrons in the wires. It is explicitly 
given by\cite{sassetti1}
\be
P(E,\{ x_{i}\})=\frac{1}{2\pi}\int_{-\infty}^{\infty} \, dE 
e^{iEt}e^{-[W(t,\{ x_{i}\})-W_{0}(t,\{ x_{i}\}]},
\ee
where $W_{0}$ is the correlation function of the corresponding 
noninteracting system. 

The connection with the standard theory of CB is established by 
observing that the steady-state current of Eq. (\ref{steady})
can be rewritten as
\beq
I&=&
\frac{8 \G_{0}^{2} }{\pi v_{F}}
\int_{0}^{\infty} dx_{1}\ldots dx_{4} \, g(x_{12}) 
g(x_{34}) \nonumber \\
&\times&
\int_{-\infty}^{\infty}dE  dE' f(E)[1-f(E')]P(E+V-E',\{ x_{i}\})
\nonumber
\eeq
where $f(E)=1-\theta(E)$ is the 
zero-temperature Fermi function. 
For noninteracting electrons the tunneling 
becomes elastic, and  
$P(E,\{ x_{i}\})=\delta(E)$ for every $\{ x_{i}\}$, thus
recovering the Ohmic $I$-$V$ curve
$I \propto V$, as it should be.
In the semiclassical approach\cite{theor1,theor2}
the function $P$ gives the probability 
of exchanging energy with the electromagnetic environment, which in 
the present microscopic theory is 
replaced by the bath of elementary excitations of the interacting 
quantum system.

The average in Eq. (\ref{steady}) can be evaluated by resorting the 
open-boundary bosonization method.\cite{obb1,obb2,obb3}
It has been shown that the low-energy properties of the isolated 
semi-infinite wire $j$ with open boundary conditions can be 
described in terms of (say) right movers only which live in an 
infinite system without boundaries,
and which are related to the left movers by the relation
\be
\psi_{j L}(x)=\psi_{j R}(-x).
\ee
In the bosonization language the above relation implies that
\be
\phi_{j L}(x)=\phi_{j R}(-x)+\mathrm{const},
\label{scalar}
\ee
where $\phi_{j \a}(x)$ is the boson field such that
\be
\psi_{j\a}(x)=\frac{1}{\sqrt{2\pi a} }
e^{-2\sqrt{\pi}\,i \e_{\a} \phi_{j\a}(x)} \, e^{i \e_{\a} k_{F}x},
\label{bospsi}
\ee
where
$k_{F}$ is the Fermi momentum and $a$ a short-distance cutoff.
The great advantage of the bosonization technique  is that the interacting 
ground-state appearing in Eq. (\ref{steady}) is nothing 
but the vacuum of the boson operators $b_{j \a q}$ entering in the 
mode expansion
\be
\phi_{j \a}(x)=i 
\e_{\a}\sum_{q>0}\frac{e^{-aq/2}}{\sqrt{2\mathcal{L}q}}
[ C_{q +} b_{j\a 
q}^{\dag}   
-  C_{q -} b_{j\bar{\a} q} ] e^{-i\e_{a}q}+ \mathrm{h.c.},
\ee
where $\mathcal{L}$ is the length of the wires.\cite{nota} The 
coefficients $C_{q \pm}$ carry all the information about the 
electron-electron interaction and are given by
\be
C_{q \pm}=\frac{1\pm K_{q}}{2\sqrt{K_{q}}},
\ee
with $K_{q}=(1+\frac{U_{q}}{\pi v_{F}})^{-1/2}$, $U_{q}$ being the 
Fourier transform of $U(|x|)$.
The special value $K_{0}\equiv K $ is the so-called LL parameter,
and, as we shall see below, it governs the power-law behavior of the 
observables within the present theory.\cite{haldane}

It is worth recalling that a single-channel conductor in 
series with a resistance $R$ can mimic\cite{safi} a LL
with an impurity (or alternatively with a tunnel junction) 
and $K=(1+R/R_{0})^{-1}$. 
Therefore our theoretical treatment could also serve to describe 
mesoscopic resistive systems, see e.g. the 
recent experiment in Ref. \onlinecite{exp12}, in which a LL with
$K\approx 1/2$ was simulated.\cite{safi}

\section{Point-like tunneling and interaction: CB regime}
\label{sec3}

In this Section we briefly review the properties 
of a junction with  point-like (edge-to-edge) 
tunneling $g(x,y)=\delta(x)\delta(y)$
and point-like (short-range) interaction $U(|x-y|)=U\delta(x-y)$
(i.e. $U_{q}=U$). In this case the function $W$ becomes particularly 
simple and it is given by
\be
W(t)=\frac{2}{K}\log \frac{a+ivt}{a},
\ee
where $v=v_{F}(1+\frac{U}{\pi v_{F}})^{1/2}$ is the renormalized 
velocity. The temporal integral in Eq. (\ref{steady}) can be 
evaluated analytically and the steady-state current reads
\be
I=\frac{8\G_{0}^{2} \,V}{\pi v^{2}\G(2/K)} 
\left(\frac{aV}{v} 
\right)^{2/K-2}.
\label{origin}
\ee
This is the well-known result originally derived by Kane and Fisher\cite{kane} by 
means of renormalization group arguments.  
For an arbitrary weak interaction the system becomes 
insulating, with the tunneling current  
suppressed (at zero temperature)
as a power-law (CB regime).
It is important to notice that, despite the power-law suppression
of the current has been observed in several experiments,
\cite{exp4,exp5,exp6,exp7,exp8,exp9,exp10,exp11,exp12}
the above formula does not recover the SO behavior 
that must hold at large bias. In the following Section
we show how this problem has been solved.

\section{Finite-range interaction and point-like tunneling: SO regime}
\label{sec4}

As anticipated in the Introduction, this case has been investigated in Ref. 
\onlinecite{sassetti1}. Here we only summarize the main conclusions, assuming 
a screened interaction of the form 
$U(|x-y|)=\frac{U}{d}e^{-|x-y|/d}$.
We stress, however, that
the results do not depend on the explicit form of the 
interaction.
The finite range of the interaction is encoded in the 
momentum-dependent functions 
\beq
K_{q}&=&\left[1+\frac{U}{\pi v_{F} (1+q^{2}d^{2})}\right]^{-1} ,
\nonumber \\
v&=&v_{F}\left[1+\frac{U}{\pi v_{F} (1+q^{2}d^{2})}\right],
\label{kq}
\eeq
as well as in the charging energy of the junction $V_{d}=2U(0)=2U/d$.
A small bias probes the low-energy (i.e. low-momentum 
$q$)  excitations of the electron liquid, and hence in 
this regime the system behaves as the interaction was 
zero-range with $K_{q} \approx K = (1+\frac{U}{\pi v_{F}})^{-1}$.
Accordingly for $V\ll V_{d}$ the same behavior $I\propto V^{2/K-1}$ as in Eq. 
(\ref{origin}) is found. 
A large bias, instead, probes high-$q$ excitations, for which 
$K_{q} \approx 1$ [see Eq. (\ref{kq})], like in the non-interacting system.
As a consequence the SO regime is correctly recovered, where the effects of
correlation manifest in a shift in the $I$-$V$ curve $I\approx 
V/R_{T}-V_{d}$, where $R_{T}=8\G_{0}^{2}/\pi 
v^{2}$ is the tunneling resistance of the 
junction. 
The crossover between the two regimes occurs at the critical voltage  
$V=V_{d}$.
We would like to mention that it has been recently shown that the 
finite-range interaction is also at the origin of the current 
suppression at small $V$ in single-channel quantum dot tunnel 
junctions, whereas a point-like $U$ produces an Ohmic behavior.\cite{perfetto2}

\begin{figure}[t]
\includegraphics[width=7.cm]{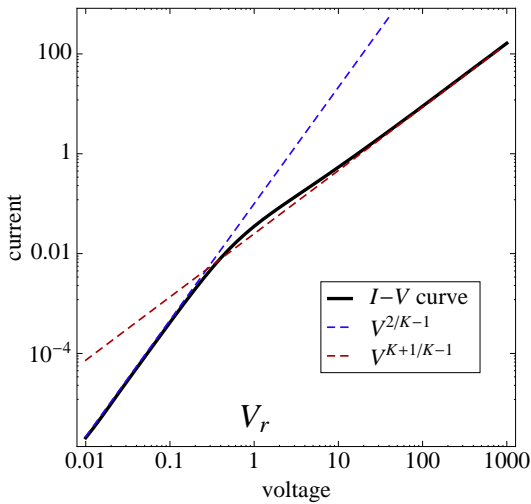}
\caption{Log-log plot of the $I$-$V$ curve for extended tunneling 
with range $r=2 \times 10^{5}a$ and 
point-like interaction with $U=6v_{F}$ (i.e. LL parameter $K \approx 0.6$).
The dashed lines represent the two power-laws with different 
exponents holding for $V<V_{r}$ and $V>V_{r}$ ($V_{r}\approx 1$ 
in this figure). Voltages $V$ and $V_{r}$ are 
in units of $10^{-5}v_{F}/a$, and the current $I$ is in units of 
$10^{-5}\G_{0}^{2}/a v_{F}$. }
\label{fig2}
\end{figure}

\section{Finite-range tunneling and point-like interaction: ET regime}
\label{sec5}

For illustration  we consider a linear junction like the one in
Fig. \ref{fig1}a. However, the explicit choice of the geometry cannot 
affect qualitatively the results.
The finite-range tunneling amplitude is in this case
$g(x,y)=e^{-(r_{0}+x+y)/r}$, where $r_{0}$ is the spatial 
separation between the edges of the wires,\cite{nota1} and $r$ is 
the size of the extended contact. We are aware that the most accurate form
of the spatial-dependent tunneling amplitude 
$g$ is probably gaussian, since it is proportional to the  overlap 
between states from the two sides of the junction.\cite{exttunn3}
Nevertheless, we here prefer to adopt the same exponential function
for both tunneling amplitude and interaction, in order make direct
comparisons (see next Section). Since we can absorb the 
factor $e^{-r_{0}/r}$ into the value of $\G_{0}$, we take 
$r_{0}=0$ without loss of generality. In this Section
we first consider a point-like 
interaction $U_{q}=U$ which makes the calculation analytically tractable,
thus allowing to get transparent formulas to  
disentangle the effects of ET.
To simplify the calculations we
approximate the integral in Eq. 
(\ref{steady}) as
\be
I \approx
\left(\frac{2 \G_{0} }{\pi a} \right)^{2}
\int_{0}^{\infty} dx  \, e^{-2x/r}  j(x) ,
\label{stadyeasy}
\ee
where $j(x)=\int_{-\infty}^{\infty}dt \,
e^{iVt}e^{-W(t,x)}$ is the steady-state current of an effective 
junction in which the tunneling occurs only  between the points at
position $x$ in both wires. 
This means that we are assuming that
the dominant contribution to the current comes from the tunneling 
events in which $x_{1}=x_{2}=x_{3}=x_{4}=x$.\cite{nota3} 
Within this approximation the function $j(x)$ can be evaluated analytically in the limits of 
small and large (compared to scale $V_{x}=v/x$) bias. 

\begin{figure}[t]
\includegraphics[width=7cm]{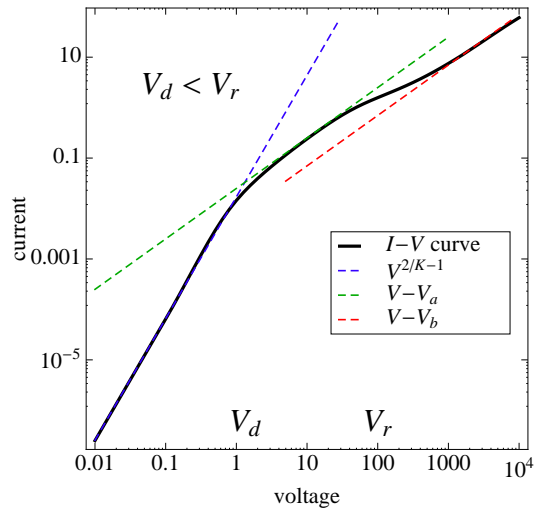}
\caption{Log-log plot of the $I$-$V$ curve for extended tunneling  
and finite-range interaction for $V_{d}<V_{r}$. We used $r=2\times 
10^{3}a$, $d=10^{6}a$, $U=6v_{F}$ (i.e. $K_{0}\approx 0.6$, 
$V_{r}\approx 100$ and $V_{d} \approx 1.2$).
The dashed lines 
represent the three power-laws with different exponents holding for 
$V<V_{r}$, $V_{r}<V<V_{d}$ and $V>V_{r}$. Voltages and current are in 
the same units as in Fig. \ref{fig2}. In the legend $V_{a}$ and 
$V_{b}$ indicate the different offset of the two shifted Ohmic regimes.}
\label{fig3}
\end{figure}

At small bias  $V \ll V_{x}$  the function $\mathrm{exp}[-W(t,x)]$ is 
dominated by the singularities around $t=\pm 2x/v$, yielding
\be
j(x) \propto x^{1/K-K} V^{2/K-1}
\quad \mathrm{for} \; V\ll V_{x}.
\ee
In the opposite limit the function $\mathrm{exp}[-W(t,x)]$ is instead 
dominated by the singularity around $t=0$, and the asymptotic current 
$j$ 
is independent on $x$,
\be
j(x) \propto  V^{K+1/K-1}
\quad \mathrm{for} \; V\gg V_{x}.
\ee
The crossover between the two regimes occurs at bias $V \approx 
V_{x}$.
In order to obtain the true tunneling current $I$,
we have to integrate $j(x)$ according to Eq. (\ref{stadyeasy}).
The numerical result is show in Fig. \ref{fig2}, where it is clearly 
seen that the crossover is also displayed by $I$ under the  
replacement   $x\to r/2 $ and $V_{x}\to V_{r}=2v/r$. 
The physical interpretation of this behavior is the following: For an 
edge-to-edge tunneling, the current is suppressed as $V^{2/K-1}$ [see 
Eq. (\ref{origin})] due to the interaction-induced depletion of density 
of states in the proximity of the boundary;\cite{obb3} allowing 
electrons to {\it tunnel over} the depletion region, {\it enhances}
the current according to a power-law with exponent $K+1/K-1 < 2/K-1$, provided 
that the energy supplied by the applied voltage is larger than the tunneling 
energy $\sim V_{r}$ (ET regime).
In the next Section we will present the most important part of the 
paper, in which we consider the simultaneous effect of extended 
contacts and screened interaction. This will allow us to study the
competition between the two energy scales $V_{d}$ and $V_{r}$, and 
see the impact on the CB scenario.

\begin{figure}[t]
\includegraphics[width=7cm]{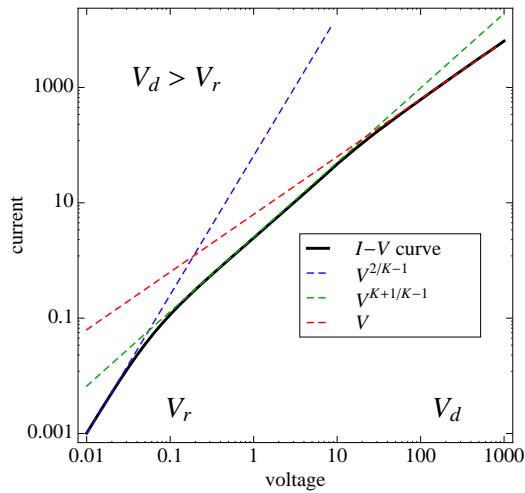}
\caption{Log-log plot of the $I$-$V$ curve for extended tunneling  
and finite-range interaction for $V_{d}>V_{r}$. We used 
$r=2\times 10^{6}a$, $d=10^{4}a$, and $U=6v_{F}$
(i.e. $K_{0}\approx 0.6$, $V_{r}\approx 0.1 $ and $V_{d} \approx 120 $). 
The dashed lines 
represent the three power-laws with different exponents holding for 
$V<V_{r}$, $V_{r}<V<V_{d}$ and $V>V_{r}$.  Voltages and current are in 
the same units as in Fig. \ref{fig2}. }
\label{fig4}
\end{figure}

\section{Finite-range tunneling and interaction: Competition}
\label{sec6}

We now consider the tunneling amplitude $g(x,y)=e^{-(x+y)/r}$
and the screened interaction
$U(|x-y|)=\frac{U}{d}e^{-|x-y|/d}$.
As noticed in Ref. \onlinecite{vignale} 
the length scales $r$ and $d$ are typically of 
the same order, and hence it is important to treat their effects on the
same footing. According to the results of the previous Sections,
if the applied bias is smaller than 
$\min \{ V_{r},V_{d} \}$ the system is certainly in the CB regime, 
with the current suppressed as  
$I\propto V^{2/K-1}$ (see  Figs. \ref{fig3} and \ref{fig4} ).
However, at larger bias the response crucially depends on the 
interplay between tunneling and screening.

\subsection{$V_{d}<V_{r}$}

If $V_{d}<V_{r}$  a SO behavior (with offset $V_{a}$, see Fig. \ref{fig3})
is expected in the range
$V_{d}<V<V_{r}$, since ET effects are still not 
significant. But what happens when $V>V_{r}$?
Tunneling effects will compete with screening effects, 
compelling the system to abandon the Ohmic behavior and to crossover towards
the power-law regime $I\sim V^{K+1/K-1}$. 
To understand the fate of such competition we have to calculate 
$I$ numerically.
For simplicity we adopt the same approximation
as in Eq. (\ref{stadyeasy}),  and the resulting  $I$-$V$  curve is 
shown in Fig. \ref{fig3}.
We see that for $V>V_{r}$ no real crossover occurs, and the current 
remains Ohmic (there is a kink separating two different SO regimes).
The physical reason of this behavior can be understood 
as follows: Since the interaction range is finite, for $V>V_{d}$ the system
behaves as it was noninteracting, where the effects of 
interaction are only visible in the Coulomb offset of 
the linear $I$-$V$ curve; as a 
consequence, when $V > V_{r}$ tunneling effects are felt by  
a ``noncorrelated state'' having $K \approx 1$ and hence 
$I \propto V^{K+1/K-1}\approx V$.  Indeed at $V \approx V_{r}$ 
a ``transition'' between two different  Ohmic regimes (characterized 
by different offests $V_{a}$ and $V_{b}$) is observed, see the green
and red dotted lines in Fig. \ref{fig3}. 
In conclusion for  $V_{d}<V_{r}$ the ET regime 
characterized by $I 
\propto V^{K+1/K-1}$ is completely suppressed.

\begin{figure}[t]
\includegraphics[width=6.2cm]{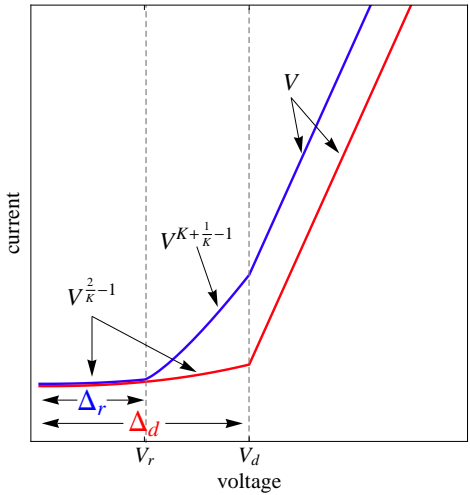}
\caption{Schematic illustration of the renormalization of the
Coulomb blockade gap $\D$. Red line: $I$-$V$ curve for a point-like
(end-to-end) tunneling and finite-range interaction, with usual gap 
$\D \approx 
V_{d} \equiv \D_{d}$.  Blue line: $I$-$V$ curve  for extended  tunneling and 
finite-range interaction with $V_{d}>V_{r}$. The Coulomb 
blockade gap renormalizes as $\D \approx V_{r} \equiv \D_{r}<\D_{d}$. }
\label{fig5}
\end{figure}

\subsection{$V_{d}>V_{r}$}

If $V_{d}>V_{r}$ the analysis is simpler, but the scenario is more 
intriguing.  In this case there is no real competition between 
ET and screening since $I 
\propto V^{K+1/K-1}$ develops in the range
$V_{r}<V<V_{d}$ while the SO behavior naturally establishes in the 
``noncorrelated'' regime at large bias $V>V_{d}$. Indeed in 
Fig. \ref{fig4} we can observe 
the three different regimes displayed by the $I$-$V$ curve which has 
been calculated 
numerically. Remarkably the occurrence of the 
ET regime {\it  before} the occurrence of the SO behavior causes a
reduction of the CB gap, because $I$ is suppressed as $V^{2/K-1}$ only for 
$V<V_{r}$ (instead of $V<V_{d}$). Since the $K+1/K-1<2/K-1$ for any 
repulsive interaction, we conclude that beyond the threshold 
$V=V_{r}$ the current is enhanced according to a ``weakly correlated''  
power-law (although the regime is still not completely Ohmic).

This finding may have relevant consequences from the experimental 
side, in particular for what concerns the estimate of the junction
capacitance $C$. Indeed $C$ is usually inferred from the relation
$1/2C=\D$, where $\D$ is the observed Coulomb blockade gap, 
identified as the high voltage offset  of the $I$-$V$ 
curve.\cite{exp4,exp9}
Our results point out that in situations in which the  
screening length is smaller than the spatial extension of the 
tunneling processes, the  relation $\D \approx V_{d}$ must be 
replaced by $\D \approx V_{r}$ (for an illustration, see Fig. \ref{fig5}).
This means that the observed gap is not simply equal to the conventional
charging energy, but it is strongly renormalized by the energy
that electrons need to tunnel over an
extended region of size $v/\D$. 

The novel regime we propose could be experimentally realized
in tunnel junctions involving multiwall carbon nanotubes. 
These systems display a manifest LL behavior\cite{exp9,egger}
and at the same time the screening by nearby gates (or substrate) and by the different 
shells renders the interaction short-ranged.\cite{egger,egger2}
Thus the condition $V_{d}>V_{r}$ can be effectively fulfilled, and
extracting the value of the capacitance from the
offset of the $I$-$V$ curve
may provide a result significantly larger than the correct one.

\section{Spinful case}
\label{sec7}

So far we have considered spinless electrons. In this Section we introduce the 
spin degrees of freedom, and show that the above scenario survives 
also in this case. 
The formulation is very similar to the one presented in Section 
\ref{sec2}.  
If the spin is taken into account the boson field $\phi_{j \a}$ 
 introduced in Eqs. (\ref{scalar},\ref{bospsi}) 
becomes explicitly spin-dependent, which we denote by $\phi_{j \a \s}$,
where $\s=\ua,\da$ is the spin orientation.
If the interaction does not depend on the spin of the scattering 
electrons [i.e. $U_{\s \s'}(x)=U(x)$] it is 
useful to introduce the charge/spin fields  $\phi_{j \a 
c/s}=(\phi_{j \a \ua} \pm \phi_{j \a \da})/\sqrt{2}$. In terms of 
these new fields the original spinful Hamiltonians separates\cite{haldane} in an
interacting part in the charge sector characterized by LL parameter
$K_{c}$ and velocity $v_{c}$ (which is 
equivalent to the one of the interacting spinless case) and a noninteracting 
part in the spin sector with $K_{s}=1$ and $v_{s}=v_{F}$. The 
calculation of the current follows the same line as above,
with the only difference that in this case the interacting 
ground-state  $| \Psi_{0} \ket$ is (in the bosonization language) the product
of the vacua of the charge and spin excitations respectively.
It is straightforward to verify that the competition between the ET 
regime and the SO regimes takes place as above, but with different 
power-law exponents: $I\propto V^{1/K_{c}}$ in the CB regime and  
$I \propto V^{(K_{c}+1/K_{c})/2}$ in the ET regime. Again it holds $1/K_{c} > 
(K_{c}+1/K_{c})/2$ (for any repulsive interaction), thus ensuring the
renormalization of the CB gap also for spinful electrons.
Finally we have verified that in this 
case the relevant energy scales are  $V_{d}=v_{c}/d$ and $V_{r}=v_{c}/r$,
i.e. both the charging energy and the ET energy depend 
only on the velocity of {\it charge} excitations,
as it should be. 


\section{Summary and conclusions}
\label{sec8}

We have investigated the zero-temperature 
nonequilibrium transport properties of a nanoscopic 
junction formed by two single-channel conductors linked by an 
extended contact. We have considered the simultaneous effect of 
finite-range electron-electron interaction and extended tunneling, by 
paying special attention to the Coulomb blockade phenomenon. Correlations have 
been included within the open-boundary Luttinger liquid theory, while tunneling 
processes have been treated to linear order in the tunneling 
Hamiltonian. Two relevant length scales enter in the problem, namely 
the screening length $d$ and the size of the extended contact $r$, 
and different scenarios have been discussed depending on their relative 
magnitude. When $d$ and $r$ are comparable a competition between 
screening and tunneling occurs, opening the possibility of 
identifying a new regime. In particular when $d<r$ a ``weakly 
correlated'' regime at intermediate voltage $V$
establishes between the well-known Coulomb 
blockade regime (holding at small $V$) and the shifted Ohmic regime 
(holding at large $V$). This produces an increase of the tunneling 
current from the CB suppression $I\sim V^{2/K-1}$ to the enhanced 
power-law $I\sim V^{K+1/K-1}$. As a consequence the CB gap shrinks 
from the ``electrostatic'' value $\D \sim 2U(0)$ to the renormalized 
value $v/r$, which is not the charging energy of the junction, but it 
is rather the energy that must be supplied to a single electron to 
tunnel over an extended region of size $r$. 
Finally we have shown that the above results are robust with respect 
to the introduction of the spin degrees of freedom, whose effect 
consists in  modification of the power-law exponents in 
the CB and ET regimes.


\begin{thebibliography}{99}
\bibitem{nota}
At the end of the calculation of any boson average
the length $\mathcal{L}$ must be sent to infinity, by remembering
that $\frac{2\pi}{\mathcal{L}} \sum_{q>0} \to \int_{0}^{\infty}dq$.
\bibitem{nota1}
If the ending points of the two wires are separated by a distance 
$r_{0}$, 
the distance between $x$ and $y$ is
$r_{0}+x+y$, since in both wires positions are positive 
definite, see Eqs. (\ref{model},\ref{tunnham}).
\bibitem{nota3}
The evaluation of the full integral $\int dx_{1} \ldots dx_{4}$ in 
Eq. (\ref{steady}) is computationally very demanding, but provides 
results in good agreement with the approximated expression in Eq. 
(\ref{stadyeasy}), due to the strong peaked shape of the correlators 
like $\bra  \psi_{j \a}(x)  \psi_{j\b} ^{\dag} (x')\ket$ around $x=x'$.
\end{thebibliography}
\end{document}